**A volumetric Modulated Arc Therapy (VMAT) combined with deep inspiration breath hold (DIBH) technique for adjuvant irradiation for left sided breast cancer.**

C.Tamburella, G.Guibert, O. Santa Cruz, P.Weber, P.Tsoutsou

*Department of Radiation Oncology, Hopital Neuchâtelois, La Chaux de Fonds, Switzerland.*

**Abstract.**

Background: The deep inspiration breath hold technique (DIBH) is widely used for left- sided breast radiotherapy (RT), in order to reduce the dose to the heart and the risk of cardiotoxicity. The volumetric arc therapy (VMAT) technique increases the dose conformity to the planning target volume (PTV). We hereby describe the procedure of combining VMAT and DIBH without a dedicated equipment, and report a dosimetric study related on our implemented VMAT-DIBH technique.

Methods: The DIBH technique is based on voluntary breath hold of the patient which is controlled by a laser on tattoo marks. Patients were selected depending their ability to breath hold for 25s, and VMAT plans were optimized with an arrangement of 4 arcs with a maximum delivery time of 20s each. A retrospective dosimetric study was undertaken on 30 patients treated with this technique: 10 of them received local (whole breast) RT (L-RT) with a simultaneous integrated boost (SIB), 10 had locoregional (LR) RT (supraclavicular, axillary, internal mammary chain regions) of the chestwall, and 10 received a LR RT of the whole breast with SIB. For these patients, their DIBH-VMAT treatment plans were compared to free breathing (FB) VMAT plans (with their same respective treatment volumes), and dose characteristics of PTV and organs at risk (OAR) were evaluated.

Results: The PTV coverage was identical among plans under DIBH or FB: 95% of the volume covered by 95% of the prescribed dose. The mean heart dose was reduced from $2.8\pm0.9$Gy in FB to $1.8\pm0.6$Gy in DIBH ($p<0.006$) for local RT, and reduced from $3.4\pm1.3$Gy in FB to $2.2\pm0.5$Gy in DIBH ($p<0.004$) for LR-RT. The ipsilateral mean lung dose was reduced by a factor of 12% ($p<0.009$) for LR-RT, but remained unchanged for L-RT with DIBH-VMAT compared to FB treatment. No significant differences were observed in the mean dose of contra-lateral organs with either DIBH or FB VMAT techniques.

Conclusion: The VMAT-DIBH technique performed without a dedicated breath hold equipment, provided important dose reductions at the heart, while preserving the ipsilateral lung and contralateral normal tissues. VMAT-DIBH might represent a clinically relevant method to reduce dose to the heart and ipsilateral lung during left breast RT.

## Introduction

Adjuvant radiotherapy (RT) plays a central role in the curative setting of breast cancer (BC) treatment [1-6]. However, RT especially when associated with systemic treatments (trastuzumab, anthracyclines, taxanes) can be associated with several long term risks of cardiotoxicity [7]. It has recently been shown that every Gy delivered to the heart is associated with a 7% increase of coronary events [8]. Therefore, the benefit of RT should be weighted against the risk of late toxicity, and methods to decrease the dose delivered to the heart are clinically relevant.

3D conformal radiotherapy (3DCRT) with tangential beams is regularly used for breast cancer treatment. New technologies, such as Volumetric Arc Therapy (VMAT) or Tomotherapy, allow the delivery of a precise "sculpted" and less heterogeneous dose to concave planning target volumes (PTV) [9]. VMAT is an emerging suitable technique for breast irradiation, allowing optimal dose sparing of surrounding organs at risks (OAR), such as the lung and heart, and being competitive with 3DCRT to avoid the contralateral breast [10-12].

The deep inspiration breath hold (DIBH) technique is well known to reduce doses delivered to the heart, by increasing the distance between the heart and the breast [13-18]. The combination of VMAT with DIBH is increasing and performed with commercially available respiration monitoring devices. The two main gated respiratory systems (GRS) used for DIBH treatments, are Real time Position Management (RPM) from Varian and the Active Breathing Coordinator (ABC) from Elekta [19-21], while several other respiratory monitoring systems (commercially available or not) are under development [22-25].However, these systems are costly and might not be widely available.

We hereby describe a low-cost technique of associated DIBH-VMAT delivered without a dedicated equipment. The method is based on voluntary breath hold of the patient which is checked by a laser on the tattoo marks throughout the treatment. This technique has already been used in other facilities, but the DIBH was combined with 3DCRT [26-27]. A multi disciplinary working group of physicians, physicists and radiation therapists (RTT), elaborated a detailed procedure before the implementation of this technique. All physicians, physicists and RTTs involved in the treatment delivery, underwent a dedicated training to the predefine procedure, before applying this method to patients. Between 2016-2018, we have treated 451 breast cancer patients with VMAT in our institution, 183 of whom with the VMAT-DIBH technique described in this publication.

In this manuscript, we describe the implementation of VMAT-DIBH, we report a dosimetric study on 30 of the above mentioned patients comparing DIBH-VMAT with free breathing (FB) VMAT (for the same patients and treatment volumes), and discuss the advantages and pitfalls of this technique.

## Materials / Methods

### A- Procedure.

**Computed Tomography (CT) simulation**

Given that no dedicated equipment was used for respiratory monitoring, the DIBH was actively monitored by the radiation therapist throughout the whole procedure. The RTT instructed the patient to breath-hold and then checked the positioning by using as reference the vertical laser. During the first breath hold, the RTT made a temporary mark, as a reference to determine whether the DIBH was reliable. For each DIBH, the laser should match this mark within a ± 3mm limit.

Before the CT simulation the RTT checked:

-   The ability of the patient to hold one's breath for more than 25s, using a manual timer.

-   The accuracy and the reproducibility of the DIBH position during the breath hold (marks on the patient).

-   The ability of the patient to breath hold appropriately at least 3 consecutive times.

A first CT scan (Philips Brillance CT 64) was performed in free breathing to determine the isocenter, which was marked on the patient's skin. Radio-opaque CT markers were then placed on this isocenter. Subsequently, a second scan was performed in breath hold position, with CT opaque markers in place, in order to determine the exact position of the isocenter in DIBH. Throughout the CT scan, the RTT actively checked, by observing the inspiration movement of the patient, that the breath hold was adequate. The two image sets (FB and DIBH) were compared after a rigid registration on Velocity (Varian). After evaluation of the distance from the heart to the thoracic wall in both CT scans, the decision to use the DIBH CT scan is taken by the physician.

**Treatment planning and quality Assurance**

Treatment plans, generated with the treatment planning system (TPS) Philips Pinnacle 14, were optimized with four 6MV photons partial arcs set as follow: first arc: from 135 to 90 ± 5 degrees CCW, second arc: from 90 to 135 ± 5 degrees CW, third arc: from 300 to 350 ± 5 degrees CW and fourth arc: from 350 to 300 ± 5 degrees CCW, in order to optimize the global treatment time. In case of a DIBH plan, a maximum duration of 13s for each single arc is introduced in the TPS Pinnacle.

The quality assurance (QA) of VMAT plans was performed by the Octavius 4D phantom (PTW Freiburg), with an absolute dose measurement. The plan acceptance was based on the gamma index pass-fail test, with the following criteria: 95% of the points had to reach the prescribed dose within a tolerance of 3%-3mm. In the process of implementation of the DIBH technique, the beam was voluntary interrupted during delivery for the first 10 patients QA, in order to check if there is any influence on the gamma index test.

**Treatment delivery**

Patient positioning was undertaken in DIBH conditions. As the gantry moved during the VMAT rotation and might hide the laser, it is necessary to use two video cameras with a good focus, in order to ensure continuous monitoring of the patient breath hold. The first camera placed on the left side of the patient, was used as the reference for the treatment position. The second one, placed at the right side, monitored the patient DIBH while the gantry hided the laser on the left side. The control of longitudinal, lateral and vertical lasers was performed every day (morning checks QA) and should remain within a tolerance of ± 2mm.

All treatments were performed on Elekta Agility 160 following the procedure described below:

For every patient treated in DIBH mode, an alert message in the treatment window of our record&verify Elekta-Mosaiq had to be approved by the treating RTT. Positioning lasers were then switched on at the beginning of the procedure. The control timer of the laser has been set for 30min which is greater than the time needed for the whole treatment. A cone beam CT (CBCT-XVI) in DIBH was performed to control the positioning of the patient before each treatment session. A fast CBCT protocol has been developed, providing an image in two acquisitions, needing only two breath holds of maximum 20s. After having applied the registration and corrections, the treatment was delivered for the 4 arcs. A radiation therapist asked the patient to breath hold and delivered the treatment when marks on the patient correspond to those set on the laser beams within a threshold of ±3mm (Figure 1). During the delivery, the RTT, trained for this technique, continuously checked if the laser remained on the skin marks. If during the treatment the patient's position moved out of these limits, the RTT immediately switched off the beam. All the registrations and corrections applied were saved in Mosaiq and for the first 30 patients, videos of the treatment were also recorded for the staff training. During the implementation phase of the technique, a properly trained physician and physicist were also present during the CBCT and treatment delivery.

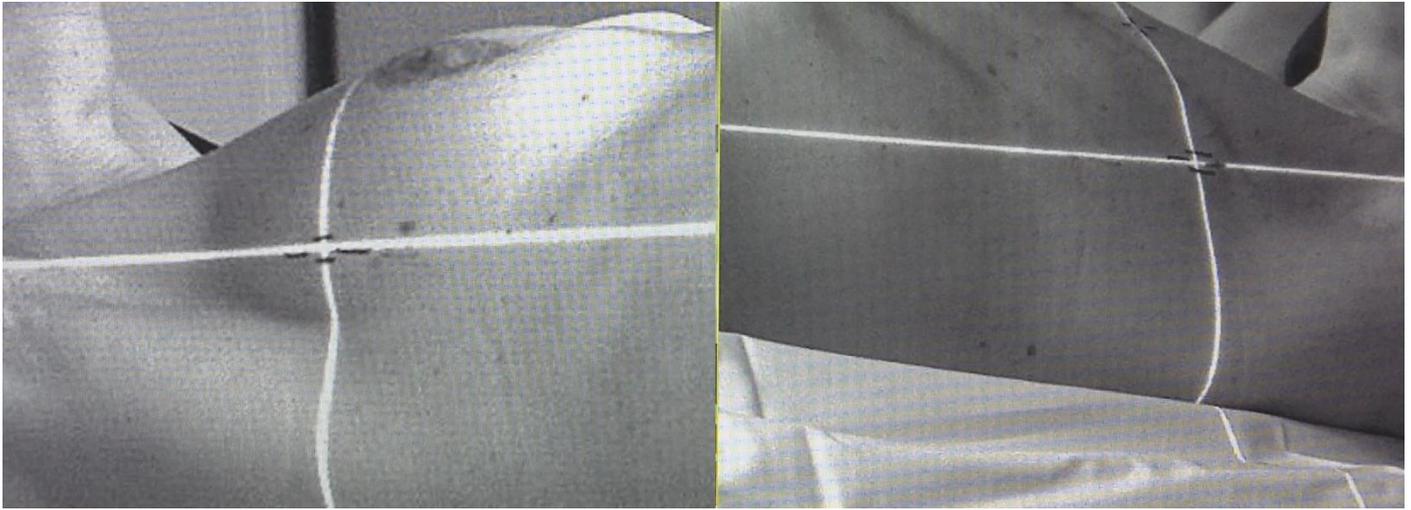

**Figure 1: Two video cameras (left &right) to check the breath hold of the patient with marks and margins of ±3mm.**

### B- Dosimetric evaluation

**Patients**

Treatment plans of 30 patients receiving adjuvant radiotherapy for non-metastatic breast cancer with the VMAT-DIBH technique were retrospectively included in this dosimetric study. Additional (non-delivered) treatment plans were calculated on the same patients and treatment volumes with FB-VMAT. The paired plans were compared for every patient.

The whole cohort was composed of 3 groups:

- 10 patients who received a local (L)-irradiation after BC surgery, with target volumes consisting of the whole left breast (50Gy; 2Gy/ fraction) with a simultaneous integrated boost (SIB) to the tumor bed (60Gy; 2.4Gy/fraction): group L-RT

- 10 patients who received a locoregional (LR)-irradiation after mastectomy, with target volumes consisting of the supraclavicular area (SCLV), the upper axilla (AX), the Internal mammary chain (IMC) (50Gy; 2Gy fraction) and the chest-wall (50Gy; 2Gy fraction) : group LR-RT ChestW

- 10 patients who received a LR-irradiation after BC surgery, with target volumes consisting of the whole breast, the SCLV and the IMC (50Gy; 2Gy fraction) with a SIB to the tumor bed: group LR-RT SIB.

**Dosimetric evaluation**

The delineation of the PTV and OAR's were performed on both scans (FB and DIBH) by the same referring physician, using the Radiation Therapy Oncology Group (RTOG) and European Society of Therapeutic Radiology (ESTRO) atlas [28,29]. For breast treatments with VMAT, it is strongly recommended to delineate the contralateral breast and lung, in order to evaluate low doses derived from beam delivery. In order to provide an accurate comparison, arc arrangement was exactly the same in FB than in DIBH, except for the delivery duration which could be increased up to 100s given that no breath hold was performed. The dose prescription, according to institutional margins, was on an extended PTV, with a margin of 5mm, to still ensure adequate PTV coverage in case of eventual misalignments. (figure 2). A virtual bolus of 1cm is systematically added to the breast for VMAT calculation in the TPS in order to simulate a bigger PTV in case of breast swell during the treatment as well as the breathing movement during FB plan delivery as properly described [30].

The dose volume histograms (DVH) of the DIBH-VMAT plans were compared to FB-VMAT plans. The plans were classified within 3 categories, corresponding to the clinical cohorts: L-RT, LR-RT-ChestW, LR-RT-SIB. A different set of normal tissue constraints was used for each context based on institutional target and normal tissue constraints (Table 1). For each patient the lung volume has been measured in Pinnacle for FB and DIBH.

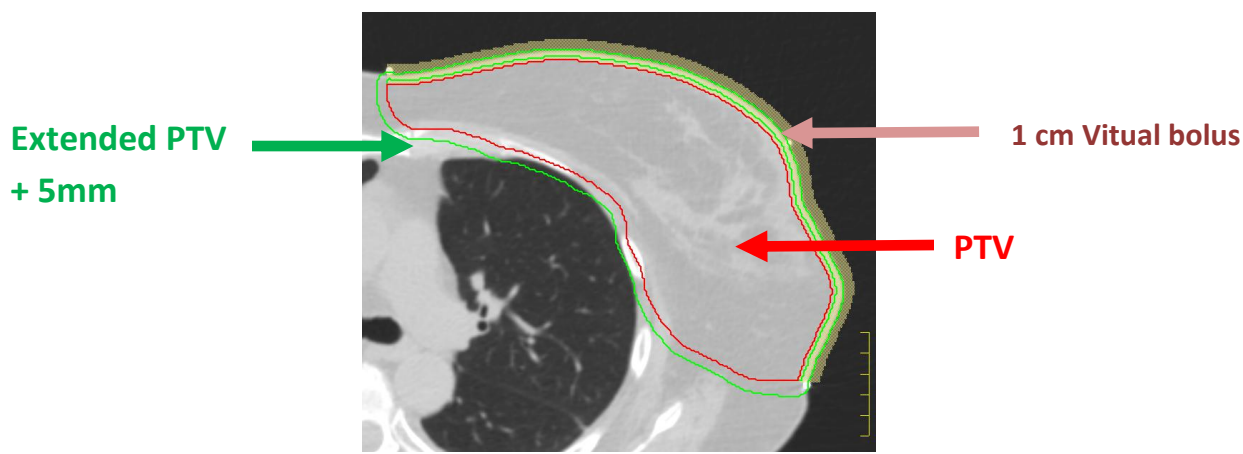

**Figure 2. PTV extended and virtual bolus.**

| Structure | Constraints | |
|---|---|---|
| | L-RT | LR- RT |
| PTV | D95% >95%<br>Dmax 115-120% | D95% >95%<br>Dmax 115-120% |
| Contra lateral Breast | V3<5%<br>Dmean< 2 (3) Gy | V3<5%<br>Dmean< 2 (3) Gy |
| Heart | V40 =0<br>V20(25)<5%;<br>V10<30%<br>Dmean <4Gy | V40<5%<br>V20<10%<br>V10<30%<br>Dmean <4Gy |
| Ipsilateral lung | V20<25%;<br>V10<35-40%<br>V5<50%<br>Dmean<10Gy | V20<35%<br>V10<35-40%<br>V5<50%<br>Dmean<10Gy |
| contralateral lung | V5<10-15% | V5<10-15% |

**Table 1: Institutional constraints of PTV and OARs for Local RT and Locoregional RT**

**Results**

### A- Procedure

The average of longitudinal, lateral, and vertical displacements needed after the CBCT image registration during the treatment for the 30 patients studied, are summarized in the table 3. These values remain within the limits of our extended PTV throughout the whole treatment for any of the given patients, maintaining an adequate PTV coverage.

All QA measurements respected the gamma index criteria with an average of 97.0±1.3% points within the tolerance of 3%-3mm. The interruption of the beam was tested during QA measurements of the first 10 treatments. At the moment the 'stop button' was activated, the beam was immediately switched off (response time <0.5s). No influence was seen on the QA procedure, by stopping the beam during the dose delivery.

During the QA procedure, the delivery time of each arc was measured with a manual timer. No discrepancy between the planned and the delivered time was noticed. A latency period of 4.0±0.5s, between the time when the 'Start' button is pressed until the exact moment the beam is turned on, has to be taken into account, which increased the total delivery time to 20.0±2s and 22.0±3s for L and LR irradiation, respectively.

No major difficulty for the patients to perform the breath hold procedure for 25s was observed. All of the treatments were successfully delivered, with no need to transform a DIBH treatment to a FB one.

### B- Dosimetric evaluation

The dosimetry data for the PTV and OAR by each group and by technique are summarized in **Table 2**.

The PTV coverage was the same for both FB and DIBH techniques for all groups and no significant differences were observed for the maximum dose. Doses received by OARs with FB-VMAT and DIBH-VMAT were compared and are highlighted in the mean DHVs by patient group (**Figure 3**).

An important reduction in the dose delivered to the heart with the DIBH-VMAT technique was achieved. The mean heart dose, the V10Gy and V20Gy were statistically significantly lower by DIBH compared to FB. The mean heart dose decreased by a factor of 40% by DIBH compared to FB (p<0.005), with a value of 1.8±0.6Gy and 2.2±0.8Gy for L-RT and LR-RT respectively. The DIBH-VMAT technique reduced the ipsilateral lung mean dose by a factor of 12% (p<0.009) for LR-RT, but remained unchanged for L-RT with DIBH-VMAT compared to FB treatment. During DIBH, the lung volume expand by a factor of 57% (mean volume 1241 ± 340 cc in FB and 2152 ± 409 cc in DIBH) which might explain the mean lung dose reduction. Contra-lateral OARs (lung and breast) did not show any statistically significant difference in terms of dose received by either technique (p>0.2)

As no delivery time limitation was introduced to optimize FB-VMAT plans, the average delivery time given by the TPS was : 31±15s/arc, 36±11s/arc and 46±12s/arc for L-RT, LR-RTchestW, LR-RT SIB respectively. This delivery time was reduced in DIBH to: 16±2s/arc and 18±3s/arc for L-RT and LR-RT (ChestW or SIB) respectively, given the introduced constraint of delivery time / arc (<20s). Due to a larger target volume to treat, the time per arc slightly increased for LR-RT and correlated to the number of delivered monitor units (MU) **(Figure 4)**.

| Structure | Parameters | VMAT-DIBH L-RT | VMAT-FB L-RT | Difference | p_value |
|---|---|---|---|---|---|
| PTV | D95%(%) | 96.8±1.4 | 96.9±1.8 | 0% | 0.60436 |
| | Dmax/Dprescr (%) | 104.4±1.2 | 103.4±0.9 | 1% | 0.06335 |
| Contra lateral Breast | Dmean(Gy) | 0.9±0.2 | 0.9±0.3 | 0% | 0.73530 |
| Heart | V40Gy(%) | <0.1 | <0.2 | - | 0.34344 |
| | V20Gy(%) | <0.4 | 1.3±1.0 | -147% | 0.03179 |
| | V10Gy(%) | <0.7 | 3.1±2.7 | -144% | 0.02120 |
| | Dmean(Gy) | 1.8±0.6 | 2.8±0.9 | -43% | 0.00537 |
| IPSILATERAL LUNG | V20Gy(%) | 9.2±4.1 | 7.9±4.2 | 15% | 0.17484 |
| | V10Gy(%) | 16.5±5.6 | 15±7.9 | 10% | 0.00173 |
| | V5Gy(%) | 28.6±8.7 | 26.6±10.9 | 7% | 0.39838 |
| | Dmean(Gy) | 6.2±1.8 | 5.7±2.1 | 8% | 0.27975 |
| CONTRA LATERAL LUNG | V5Gy(%) | <0.1 | <0.1 | - | 0.34344 |
| Structure | Parameters | VMAT-DIBH LR-RT SIB | VMAT-FB LR-RT SIB | Difference | p_value |
| PTV | D95%(%) | 95.7±1.5 | 96.3±2.5 | -1% | 0.35716 |
| | Dmax/Dprescr (%) | 107.2±3.4 | 105.5±2.0 | 2% | 0.09113 |
| Contra lateral Breast | Dmean(Gy) | 1.9±0.7 | 1.9±0.8 | 0% | 0.53635 |
| Heart | V40Gy(%) | <0.1 | <0.1 | - | 0.34344 |
| | V20Gy(%) | <0.5 | 1.4±1.2 | -75% | 0.02386 |
| | V10Gy(%) | 1.3±1.2 | 2.8±2.6 | -73% | 0.00208 |
| | Dmean(Gy) | 2.2±0.8 | 3.4±1.3 | -43% | 0.00003 |
| IPSILATERAL LUNG | V20Gy(%) | 18.3±3.7 | 21.1±4.5 | -14% | 0.02050 |
| | V10Gy(%) | 32.3±5.4 | 35±5.9 | -8% | 0.16687 |
| | V5Gy(%) | 48.2±5.4 | 51.7±6.8 | -7% | 0.14234 |
| | Dmean(Gy) | 10.5±1.5 | 11.8±1.7 | -12% | 0.00881 |
| CONTRA LATERAL LUNG | V5Gy(%) | <0.2 | <0.3 | 67% | 0.34344 |
| Structure | Parameters | VMAT-DIBH LR-RT ChestW | VMAT-FB LR-RT ChestW | Difference | p_value |
| PTV | D95%(%) | 94.2±2.0 | 94.1±2.7 | 0% | 0.84023 |
| | Dmax/Dprescr (%) | 113.9±3.6 | 115.6±4.7 | -1% | 0.37058 |
| Contra lateral Breast | Dmean(Gy) | 2.3±1.1 | 2.3±1.0 | 0% | 0.94382 |
| Heart | V40Gy(%) | <0.1 | <0.1 | - | 0.34344 |
| | V20Gy(%) | <0.7 | 1.6±1.3 | -137% | 0.00190 |
| | V10Gy(%) | 1.7±1.4 | 3.8±3.3 | -76% | 0.00315 |
| | Dmean(Gy) | 2.2±0.8 | 3.5±1.5 | -46% | 0.00041 |
| IPSILATERAL LUNG | V20Gy(%) | 17.4±4.1 | 21±4.1 | -19% | 0.00173 |
| | V10Gy(%) | 30.6±5.3 | 34.9±5.0 | -13% | 0.00080 |
| | V5Gy(%) | 45.8±6.7 | 49.7±6.7 | -8% | 0.04298 |
| | Dmean(Gy) | 9.86±1.5 | 11.5±1.4 | -15% | 0.00020 |
| CONTRA LATERAL LUNG | V5Gy(%) | <0.6 | <0.3 | 143% | 0.21068 |

Table 2: Summary of dosimetry data for the PTV and OARs by FB-VMAT and DIBH-VMAT technique for L-RT, LR-RTChestW, LR-RT SIB groups and their respective standard deviations.

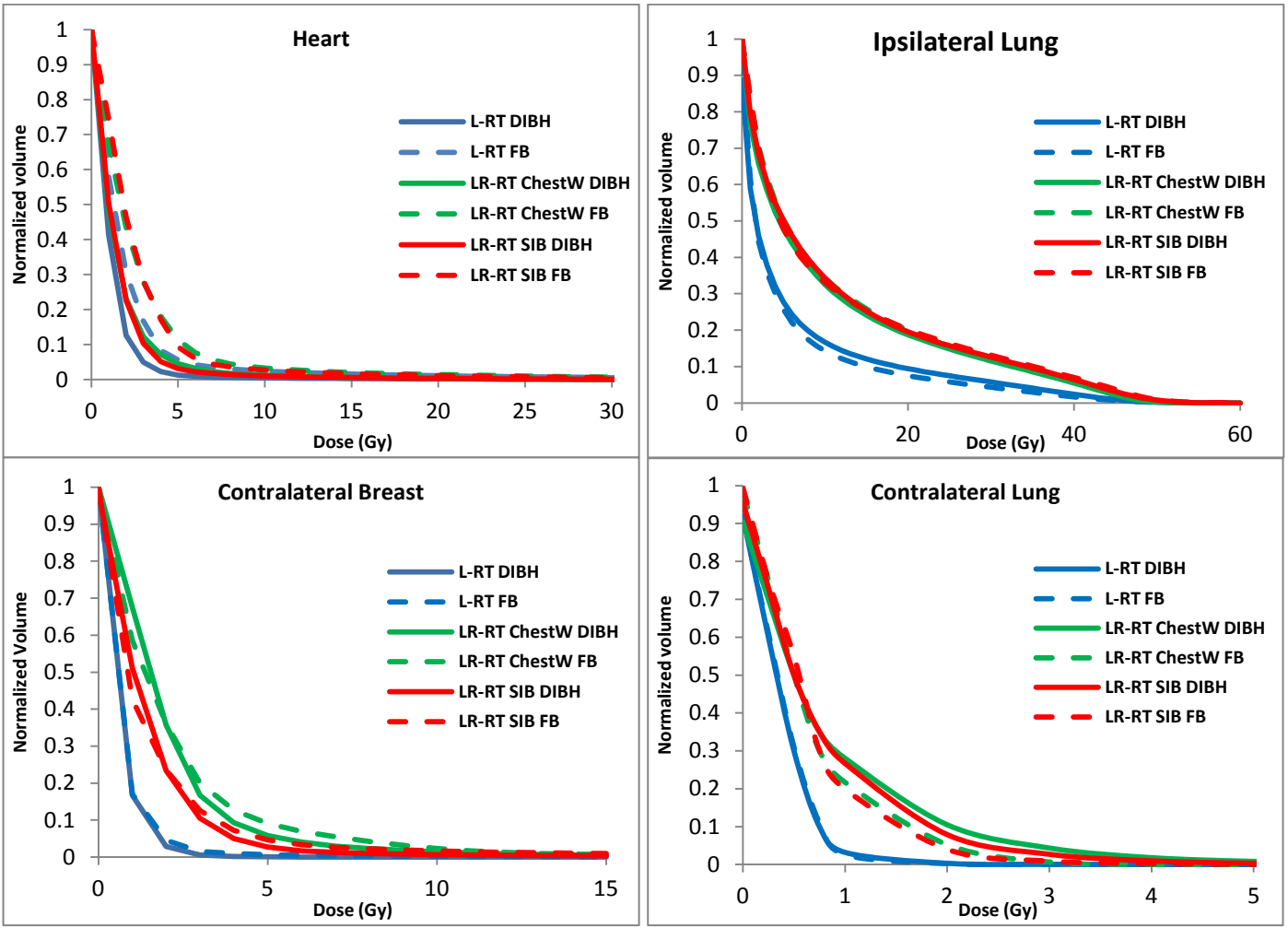

Figure 3: Mean DVHs for OARs

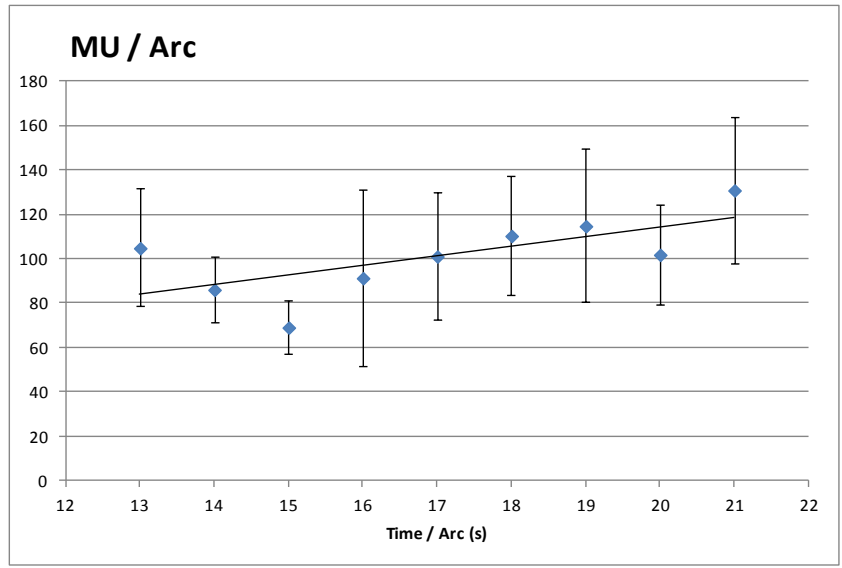

Figure 4: Correlation between the number of monitor units and the time per arc in DIBH plans

| VMAT-DIBH Displacements | Longitudinal (cm) | Lateral (cm) | Vertical (cm) |
|---|---|---|---|
| L-RT | 0.26 ± 0.16 | 0.15 ± 0.11 | 0.17 ± 0.13 |
| LR-RT SIB | 0.26 ± 0.25 | 0.27 ± 0.25 | 0.33 ± 0.23 |
| LR-RT ChestW | 0.20 ± 0.16 | 0.21 ± 0.09 | 0.17 ± 0.12 |

Table 3: Mean displacements for treatments in DIBH with their standard deviation.

## Discussion

DIBH is a well established efficient technique permitting to spare the heart from irradiation dose during left-breast radiotherapy. The DIBH technique has been developed using devoted gating respiratory systems (GRS) [31-32], and a reduction of the dose delivered to the heart has been documented. It has been also developed without a dedicated GRS during conventional 3D- breast RT [26]. The introduction of intensity modulated RT (IMRT) in breast cancer RT has shown advantages in terms of dose homogeneity and decrease in acute toxicities [33]. VMAT represents an IMRT technique that has recently been shown to offer dosimetric advantages in breast RT [11]. The association of DIBH-VMAT with a dedicated GRS has been previously reported [31]. Several studies [34-36] have demonstrated that using partial arcs is more effective in reducing low doses at the contra lateral OARs. In this context, the VMAT technique can offer several advantages compared to 3DCRT. The feasibility of fast VMAT arc delivery using the TPS Eclipse (Varian) has also been previously studied [37]. To the best of our knowledge, no study has reported so far an association of VMAT and DIBH without the use of a respiratory gating system.
Nevertheless in our study, in the case of LR-RTChestW, the dose to the contralateral breast was slightly increased by using DIBH compared to FB, although it still remained below the pre-established dose constraints. Chestwall irradiation is known to be challenging, due to the postoperative anatomy, which makes it more difficult to avoid the contralateral breast.

We aimed to evaluate a technique combining the advantages of DIBH (sparing the heart) and VMAT planification (precise and homogeneous dose at PTV) without the use of a dedicated equipment. We obtained reduced heart and ipsilateral lung doses with this technique that are comparable to those reported in the literature with a dedicated GRS system [22;24;29;31]. Of note, in our study, the ipsilateral lung dose reduction was only observed for LR-RT. These dosimetric advantages are anticipated to correlate to clinical benefits, given the well established doses of cardio-toxicity [7,8] and lung toxicity [7].

The technique we hereby describe was easy to implement and required no dedicated equipment. The reproducibility of the breath hold was very high, as pointed out by the minimal displacements during treatment delivery, although a limitation of this technique remains the fact that no documentation of the DIBH treatment could be recorded. An advantage of a dedicated GRS system is the documentation of the delivery of the treatment with DIBH. A further implementation for this technique would be: real-time imaging with the portal imager during the actual treatment delivery, or alternatively, a record of the video for each patient to document proper delivery.

Another limitation for the implementation of this technique is that it is patient- and operator- dependent and therefore relies on a strict clinical protocol. The patient should integrate clear instructions from the physician and the RTT during CT simulation. DIBH-VMAT became then technically feasible to implement, as well as cost-effective, given that only 2 video cameras were needed as extra equipment.

Patient's positioning was a key step in the procedure of DIBH-VMAT delivery without any GRS. A daily CBCT positioning imaging was necessary before each treatment to provide accuracy in irradiation delivery. After the registration and corrections applied, the whole treatment should be delivered within the 3mm tolerance, which implied an active, continuous monitoring of the video camera by the operating RTT, throughout the beam delivery. Lasers set on the skin marks were very reliable. Still the RTTs must remain active, and switch off the beam, if necessary. Therefore, a dedicated training was a prerequisite for the implementation of the technique.

Overall, this technique has been robustly implemented in our Department with more than 400 patients treated todate. Its dosimetric advantages support its use in clinical practice despite the limitations in its implementation and documentations and make it an interesting options for centers not disposing of dedicated GRS equipments.

## Conclusions

A technique combining DIBH-VMAT has been successfully implemented in our Institution, without a dedicated breath hold equipment. The delivery of the treatment was achieved in a reasonably short time. An adequate PTV coverage was obtained. The mean heart dose was significantly reduced, without any increase of the dose to the ipsilateral lung, while appropriately sparing the contralateral OARs. Sparing the heart to reduce risk of late radiation-induced toxicity is pivotal, especially for a population of patients with a long anticipated survival. We believe that this technique could be an interesting option for institutions not disposing of dedicated breath hold equipments.